\documentstyle[12pt,epsf]{article}

\newlength{\dinwidth}
\newlength{\dinmargin}
\begin{document}

\newcommand{\be}{\begin{equation}}
\newcommand{\ee}{\end{equation}}
\newcommand{\ber}{\begin{eqnarray}}
\newcommand{\eer}{\end{eqnarray}}
\newcommand{\lp}{\left(}
\newcommand{\rp}{\right)}
\newcommand{\lk}{\left\{}
\newcommand{\rk}{\right\}}
\newcommand{\lc}{\left[}
\newcommand{\rc}{\right]}
\newcommand{\ls}{\alpha'}
\newcommand{\cm}{\hspace{1cm}}
\newcommand{\r}{{\bb R}}

\baselineskip18pt

\thispagestyle{empty}

\begin{flushright}
\begin{tabular}{l}
FFUOV-99/12\\
{\tt hep-th/9907083}

\end{tabular}
\end{flushright}

\vspace*{2cm}

{\vbox{\centerline{{\Large{\bf A Matrix Model for Type 0 Strings
}}}}}

\vskip30pt

\centerline{Jes\'{u}s Puente Pe\~{n}alba
\footnote{E-mail address: jesus@string1.ciencias.uniovi.es}}

\vskip6pt
\centerline{{\it Depto. de F\'{\i}sica, Universidad de Oviedo}}
\centerline{{\it Avda. Calvo Sotelo 18}}
\centerline{{\it E-33007 Oviedo, Asturias, Spain}}

\vskip .25in

\vspace{1cm}
{\vbox{\centerline{{\bf ABSTRACT }}}}

   A matrix model for type 0 strings is proposed. It consists in making a
non-supersymmetric orbifold projection in the Yang-Mills theory and identifying
the infrared configurations of the system at infinite coupling with strings.
The correct partition function is calculated. Also, the usual spectrum of 
branes is found.
Both type A and B models are constructed. The model in a torus contains all
the degrees of freedom and interpolates between the four string theories
(IIA, IIB, 0A, 0B) and the M theory as different limits are taken.

\vspace*{24pt}

\setcounter{page}{0}
\setcounter{footnote}{0}

\newpage

\section{Introduction}

   Matrix theory \cite{bfss} has proven to be a useful tool to put together 
all the properties of the different known corners of M theory. All the 
dualities arise very naturally and non-perturbative objects 
can be handled in a way that is more dynamical. All objects with different 
dimensionalities coexist in the same Yang-Mills action and therefore the 
``brane democracy'' has a clear meaning.

   Until recently, bosonic string theories were not part of the web of 
dualities that gathered all others as realizations of the same M theory in 
different backgrounds. This theories were developed in refs.\cite{var} and 
others. In this work, I shall concentrate in type 0A and type 0B bosonic 
string theories. The connection to the supersymmetric theories is got with an 
orbifold compactification of IIA \cite{rohm} or M theory \cite{gab} that 
describes these models in the appropriate limit of certain radius. There are 
also interpolating models constructed in \cite{blum} that relate the bosonic 
and the supersymmetric strings in a continuous way. However, there are still 
doubts on the validity of the interpolating model when the coupling is not 
zero and the 
present knowledge of the branes of the theory is not complete yet. There are, 
therefore, several reasons to try to make a matrix model of these strings and 
see how much information can be obtained.

  One of the problems that can be solved with this matrix model is the 
behaviour when the coupling is not zero and, in particular, the way the
tachyon acquires a mass and ceases to yield the theory inconsistent. Also,
the duality relations are clearer because they explicitly appear in the
Lagrangian. The
introduction of D-brane backgrounds is more problematic. They appear as 
objects of the theory but their interactions and, in particular, the couplings
to the closed string fields still require a better knowledge of open type 0
strings.

\section{Description of the orbifold}

   The idea is to try to make an orbifold of the Matrix Model of M theory
so that we obtain a non-perturbative, matrix model of type 0 strings.
As argued in \cite{gab} these strings are compactifications of M
theory in an orbifolded circle. The first step is, therefore, to
compactify the model in a circle through the construction of W.
Taylor \cite{taylor} and arrive at the Hamiltonian 
\ber
H=\frac{1}{2}\frac{R_+}{ 2 \pi} \int_0^{2 \pi}\frac{d \sigma_9}{R_9}
tr \lk p_i^2+R_9^2 (DX_i)^2+ R_9 \theta^T D \theta+\right. \nonumber \\
+\left. \frac{1}{R_9^3}\lp E^2 +\lc X^i,X^j \rc^2 \rp +
\frac{1}{R_9^{3/2}}\theta^T \gamma_i \lc X^i,\theta \rc \rk 
\eer
where everything is measured in units of $l_p^3 R_9^{-1}$.
The question is now how to make a world-line orbifold of the Yang-Mills
theory such that its effects are the same as the target orbifold of
ref.\cite{gab}. As an ansatz, we shall work with a similar one and test 
its properties (spectrum, partition function, ...) to see if it, indeed, 
corresponds to a matrix realization of type 0 strings. The ansatz is to 
project out all bosons with odd world-line momentum. This way, world-line 
fermions effectively live in a circle of double size that of the bosons. Other
way of looking at it is to consider that both live in the same circle but with
different boundary conditions: bosons are periodic while fermions can be both
periodic and anti-periodic. The conditions are similar to those one imposes 
around the compact Euclidean temporal direction when calculating at finite 
temperature. The difference is that fermions can also be periodic.

   It is clear that these conditions break supersymmetry because they treat 
fermions and bosons in a different manner. One important consequence is that
the zero modes of the oscillators do not cancel. Moreover, because of the fact
that there are twice as many fermionic as bosonic modes, the vacuum has an 
infinite negative energy signalling the existence of a tachyon in the target 
theory.

   In order to properly identify the string states, it is necessary to use
other parameters
\be
\ls=\frac{l_p^3}{R_9} \hspace{1cm} g_s=\frac{R_9}{\sqrt{\ls}}
\ee
Now we can write the matrix string Hamiltonian:

\be
H=\frac{1}{2}\frac{R_+}{ 2 \pi} \int_0^{2 \pi} d \sigma_9
tr \lk p_i^2+(DX_i)^2+ \theta^T D \theta+\frac{1}{g_s^2}\lp E^2 -\lc
X^i,X^j \rc^2 \rp + \frac{1}{g_s}\theta^T 
\gamma_i \lc X^i,\theta \rc \rk
\label{hamil}
\ee

The perturbative string limit is the strong coupling limit of the Yang-Mills
theory. To analyse the spectrum we have to go to the infrared limit. There,
the scalars take vacuum expectation values and break the gauge symmetry to
its Cartan sub-algebra ($U(1)^N$). This gives masses to all the fields so that
in the extreme infrared, everything can be considered decoupled but the photons
and the scalars that correspond to the commuting part of the matrices. This 
part will give us the free strings. There are eight scalars related by the 
$SO(8)$ R-symmetry, one gauge field and their fermionic partners.

   The spectrum of these fields can be obtained expanding them in a Fourier 
series. The Hamiltonian is then a sum of fermionic and bosonic oscillators
like
\be
\frac{4 \pi}{R_+}H=\sum_n \Pi_n^2+ X_{-n} n^2 X_n + \mbox{fermions}
\ee

   It is not immediate to see the different sectors (NS-NS, R-R,...) this way 
because the action we are dealing with is the equivalent of the Green-Schwarz 
action, not the Neveu-Schwarz-Ramond version. In the latter formalism, one can 
classify the fields according to the boundary conditions of the fermionic 
operator that is used to create them from the vacuum. With the G-S action, one
just generates the whole spectrum acting with the supercharges over the vacuum
state. This is impossible in this case because we do not have any. What we 
shall do is to compute the whole spectrum and see how it is organised in terms
of the characters of the spin group.

\section{The partition function.}

  The best way of obtaining the spectrum is to calculate the internal partition
function of the theory. Usual light-cone type 0 string theory is supposed to 
emerge in the large-N limit of the gauge theory. In the limit of infinite 
coupling, the Hamiltonian is truncated to the free part. Besides, we can 
identify 
\be
\frac{N}{R_+}=p_+
\ee
Apart from the $N$ diagonal fields of each class, there can also exist
excitations with twisted boundary conditions like these
\be
X(2\pi)={\cal U}^\dagger X(0) {\cal U}
\ee
where ${\cal U}$ is an element of the $Z^N$ group of permutations. This group
is the centre of the $U(N)$ gauge group and it is not broken even when the
scalars take vacuum expectation values. For 
each element of the group there is a partition of the whole matrix in twisted
states that represent single string states whose light-cone momentum is
$p_+=N_i/R_+$. The presence of states with several strings is a
consequence of the fact that matrix models automatically include the second
quantisation of the objects they describe.  The partition function of the
usual (non-matrix) strings calculated computing the path integral of the 
world-sheet fields is, in fact, a single string partition function. As we
now want to compare the matrix results with the usual ones, we have to 
restrict the computation to single string states. With that, the final action
is the Green-Schwarz light-cone action modified so that the fermionic fields
can also be anti-periodic when going around the world-sheet coordinate.
 
  We are now ready to find out the contribution of each field. It is easier 
if we imagine that all fields live in the bosonic circle 
(radius $R_9$) and let bosons be periodic and fermions both periodic and 
anti-periodic. The partition function can be seen as a trace over states that 
can be performed calculating the propagator of the fields with periodic (or
anti-periodic) boundary conditions. The part of each boson is

\be
Z_B=\frac{1}{\sqrt{\tau_2}}\frac{1}{\eta(\tau)\bar{\eta}(\tau)}
\ee

This is obtained with the usual techniques of determinant regularization
and Schwinger integral representations. It is quite standard, so I shall not 
explain every step. The fermions can be organised in spin structures 
according to the boundary conditions one chooses in both directions. One has 
to calculate the determinant of the Dirac operator on the torus (of complex 
parameter $\tau$) with eigenvalues $(n \tau+ m)$ with $n$ an $m$ integers if 
the fields are periodic and integers plus one half if they are anti periodic. 
After a proper regularization the results are

\ber
Z_{pp}& = &\frac{\vartheta_1\lp \tau\rp}{\eta\lp  \tau\rp} \hspace{1cm}
{ \mbox{ if they are periodic-periodic }  } \nonumber \\
Z_{pa}& = &\frac{\vartheta_2\lp \tau\rp}{\eta\lp \tau\rp} \hspace{1cm}
{ \mbox{ if they are periodic-antiperiodic }  } \nonumber \\
Z_{ap}& = &\frac{\vartheta_3\lp \tau\rp}{\eta\lp \tau\rp} \hspace{1cm}
{ \mbox {if they are antiperiodic-periodic  } } \nonumber \\
Z_{aa}& = &\frac{\vartheta_4\lp \tau\rp}{\eta\lp \tau\rp} \hspace{1cm}
{ \mbox {if they are antiperiodic-antiperiodic}   } 
\eer
for the right moving part of each field. 
The first condition refers to the $\tau$ direction. After multiplying the
bosonic and the fermionic left and right-moving parts and summing over all the
possible spin structures, the result is
\be
Z=\frac{1}{2\tau_2^4}  \left| \eta(\tau)\right|^{-24} \lp 
\left| \vartheta_2(\tau)\right|^8+\left|\vartheta_3(\tau)\right|^8+\left|
\vartheta_4(\tau)\right|^8 \rp
\ee

 This is exactly the partition function of type 0 string theories. In our case
the chirality of fermions is that of an A theory so this
should be a description of type 0A string theory. 
  With this function we can analyse the sectors of the spectrum reordering
the Jacobi functions. One can put it this way
\be
Z=\frac{1}{4 \tau_2^4}  \left| \eta(\tau)\right|^{-16} \lc 
\frac{ \left|\vartheta_2(\tau)\right|^8}{\left| \eta(\tau)\right|^8}+
\frac{ \left|\vartheta_2(\tau)\right|^8}{\left| \eta(\tau)\right|^8}+
\lp \frac{ \vartheta_3^4(\tau)}{\left| \eta(\tau)\right|^4} -
\frac{\vartheta_4^4(\tau)}{\left| \eta(\tau)\right|^4}\rp^2+
\lp \frac{\vartheta_3^4(\tau)}{\left| \eta(\tau)\right|^4} +
\frac{\vartheta_4^4(\tau)}{\left| \eta(\tau)\right|^4}\rp^2 \rc
\ee
which is

\be
Z=Z_b \lp \bar{\chi_S}\chi_C+
\bar{\chi_C}\chi_S +\bar{\chi_V}\chi_V+\bar{\chi_I}\chi_I\rp
\ee
where the $\chi$ functions are the characters of the different representations
of the target Lorentz group. Notice that all of them are bosonic.
The characters tell us the
rotational properties of each field. The $\bar{V}V$ sector is the 
gravitational one, it corresponds to the NS-NS sector of the supersymmetric 
theories and contains, among other fields, the graviton. The $\bar{I}I$ sector
contains the tachyon and the two others are the two R-R sectors. 
We can now find more explicitly the mass spectrum
and the degeneracies expanding the partition function in a series of 
exponentials of $\tau_2$, once the integral over $\tau_1$ is done.
 The first orders are
\be
Z=\tau_2^{-4} \lp q^{-2}+ 192 +1296 q^2 +{\cal O}(q^4)\rp
\ee
where $q=e^{-2 \pi \tau_2}$. The first addend is divergent in the infrared 
(when $\tau_2$ is large) and is the contribution of the tachyon. The second
addend corresponds to the massless fields ($64$ degrees of freedom come from 
each R-R sector and $64$ more from the gravitational one). 
The partition function is a 
modular form, invariant, in particular under the inversion of $\tau_2$. This
means that its ultraviolet behaviour is exactly the same as the infrared one.
The infrared divergence of the tachyon is the same as the ultraviolet 
Hagedorn growth of the degeneracies. If the latter requires a small temperature
to be regularized, the former would require a high one, so that finally, the
theory is not sensible at any temperature; at least with $g=0$.

\section{The long-range potential}

  One of the effects of the appearance of a tachyon is that it is impossible
to calculate a potential because the interchange of tachyons is extremely 
divergent in the infrared. Let us see that this is indeed our case. At one 
loop, it is not difficult to obtain the vacuum energy of the free fields in
the compactified theory. The easiest way is to use the background gauge
and count the number of fermionic and bosonic fields at each mass level. 
In the Schwinger representation the result is
\ber
V(r)=\frac{v}{4}\sqrt{\frac{R_+}{g_s}}\sum_n \int ds s^{-1/2} 
e^{-sr^2R_+/g_s} \frac{1}{\sinh s v}\times \nonumber \\
\lc 16 e^{-s (2\pi n)^2 g_s R_+} \cosh sv- 
e^{-s (\pi n)^2 g_s R_+} \lp 4\cosh 2 s v-12\rp \rc
\eer
when the coupling is small, one has to make a Poisson resummation and get

\ber
V(r)=\frac{v}{4}\frac{R_+}{\sqrt{\pi}g_s}\sum_n \int ds s^{-1} 
e^{-sr^2R_+/g_s} \frac{1}{\sinh s v}\times \nonumber \\
\lc 8  e^{-s n^2 /(4 g_s R_+)} \cosh sv- 
e^{-s n^2 /g_s R_+) } \lp 4\cosh 2 s v-12\rp \rc
\eer
which in the infrared behaves as
\be
-\frac{2}{\sqrt{\pi} g_s} \int \frac{ds}{s^2} e^{-s r^2 R_+/g_s}
\ee
This is divergent and denotes the existence of the tachyon in the spectrum.
It is possible for us to do perturbative computations in spite of dealing with
a strongly-coupled theory because, in fact, we have deduced the form of the
effective theory at low energies and it has turned out to be free.
It is this free theory the one we have calculated the vacuum energy for and
not the whole one. An analogous computation to this one was done in 
ref.\cite{tse}. There, the potential between two D3-branes was calculated.
It was also divergent in a similar way and was given the same 
interpretation as a tachyonic effect. In order to give a sensible result, 
the authors of that reference introduced an infrared (in the image of the
closed strings) cut-off and regularized the potential.

\section{0B strings and the dualities.}

  We shall now be interested in describing models for 0B strings and how they
can interpolate between them and the supersymmetric theories. For that, it is
necessary to compactify the theory further in a second circle (third, taking
into account also the light-like one) of radius $R_8$. After using again the 
Yang-Mills duality described in \cite{taylor} we have a theory in $2+1$ 
dimensions. Around the second circle the fields have the usual periodic 
boundary conditions. This model interpolates between all five theories (IIA,
IIB, 0A, 0B and M). If one chooses both radii to be small, the resulting 
infrared limit describes a type B theory. Which one depends on the particular 
choice of the string coupling constant
\be
g_s^{0B}=\frac{R_9}{R_8}
\ee
or
\be
g_s^{IIB}=\frac{R_8}{R_9}
\ee
In this latter case it is the coupling constant of type IIB theory 
compactified in the orbifolded circle. When we take that circle to infinity, 
it will coincide with the supersymmetric case. The Hamiltonian is

\ber
H=\frac{1}{2}\frac{R_+}{ 2 \pi} \oint d^2
\sigma tr \frac{1}{R_9 R_8} \lc p_i^2 +(R_9 R_8 \partial_0 A_8)^2+
(R_9 R_8\partial_0 A_9)^2+  R_9^2 (\partial_1 X_i)^2 +  
\right. \nonumber \\  \left.
+R_8^2 (\partial_2 X_i)^2 + 
(R_9 \partial_1 A_8 -  R_8 \partial_2 A_9)^2 + 
\theta^T ( R_9 \partial_1 +R_8 \partial_2) \theta + \mbox{interactions} \rc
\eer

with the constraint that bosons are periodic around both directions but 
fermions are periodic around $\sigma_8$ and can be periodic or anti-periodic
around $\sigma_9$. This difference between the two directions spoils the 
$SL(2,Z)$ symmetry of IIB strings. The self-duality is reduced to an S-duality
between two different theories. 

  If we choose $R_9$ to be the quantum circle. We have to change variables
this way
\ber                                                      
H=\frac{1}{2}\frac{R_+}{ 2 \pi} \int_0^{2\pi} d\sigma_9 \int_0^{2\pi/R_8}
d\sigma_8 tr \frac{1}{R_8} \lc p_i^2 +(R_9 R_8 \partial_0 A_8)^2+
(R_9 R_8 \partial_0 A_9)^2+  R_8^2 (\partial_1 X_i)^2 +  
\right. \nonumber \\  \left.
+(\partial_2 X_i)^2 + 
(R_8 \partial_1 A_9 -  \partial_2 A_8)^2 +           
\theta^T ( R_8 \partial_1 + \partial_2) \theta + \mbox{interactions} \rc
\eer

   As in the previous case, the limit where the strings give a good 
description is $g_s^{0B}\rightarrow 0$, which is strongly coupled in the
Yang-Mills. The extreme infrared limit is again described by the commuting
fields and the structure of twisted states is also the same as that in the
IIA (or 0A) theories (see \cite{sfm,motl} where this aspect, and others, are
studied for type IIA and IIB matrix string theories).

 We can now analyse the spectrum. That of the scalars is

\ber
E^2_{\mbox{scalars}}=\sum_{i=1}^7 \frac{1}{2p_+}N_{n,m}^i \frac{1}{R_9} 
\sqrt{n^2 R_9^2 + m^2 R^8}=\sum_{i=1}^7 
\frac{1}{2p_+}N_{n,m}^i \sqrt{n^2 +m^2 \frac{R_8^2}{R_9^2}}= \nonumber \\
=\sum_{i=1}^7  \frac{1}{2p_+}N_{n,m}^i \sqrt{n^2 +\frac{m^2}{(g_s^{0B})^2}}
\label{d-strings}
\eer
with both $n$ and $m$ integers. We can see the IIB-like structure of 
$(p,q)$-strings. In fact there is no difference in the bosonic sector. The 
gauge field has the same spectrum as the scalars, but one has to consider also
the purely topological sectors.
There are three different elements of $F^{\mu \nu}$. The two electric 
components have configurations like
\ber
\frac{1}{(2\pi)^2}\oint d^2 \sigma \partial_0 A_8=\frac{k}{R_8} \nonumber \\
\frac{1}{(2\pi)^2}\oint d^2 \sigma \partial_0 A_9=\frac{l}{R_9}
\eer
With this we have to add
\be
E^2_{\mbox{winding}}=\frac{1}{2p_+}\lp \frac{k^2}{R_8^2}+\frac{l^2}{R_9^2} 
\rp=\frac{1}{2p_+}\tilde{R}_8^2\lp k^2+\frac{l^2}{(g_s^{0B})^2} \rp
\ee
to the spectrum. One has to impose the condition of transversality of the 
electric field and the momentum
\be
\vec{p}\vec{E}=0
\ee
which means that
\be
a (k,l)= b (-n,m), 
\ee
where all numbers are integer. As a result, $b$ is the winding number, $a$
is the excitation number and the true tension is
$a^{-1}(n,m)=b^{-1}(k,l)=(p,q)$.
A similar quantisation condition holds for the magnetic field. In this
case the easiest way to obtain the energies is to follow \cite{bfm} and to
quantise the momentum of the string along the eighth direction. Adding up
every term, we have obtained
\ber
E^2=\frac{1}{2p_+}\lk \sum_{i=1}^8 \lc N_{n_i}^{(p,q)} n_i^{(p,q)} 
\sqrt{p^2+\frac{q^2}{(g_s^{0B})^2}}+
\omega^2 \tilde{R}_8^2\lp p^2+\frac{q^2}{(g_s^{0B})^2} \rp+ (p^i)^2 \rc +
\frac{k^2}{\tilde{R}_8^2} \rk= 
\nonumber \\
= \frac{1}{2p_+}\lc \sum_{i=1}^8 \lp N_{n_i}^{(p,q)} n_i^{(p,q)} 
2 \pi T_{(p,q)} + (2 \pi \omega \tilde{R}_8)^2 T_{(p,q)}^2+ (p^i)^2 \rp+ 
\frac{k^2}{\tilde{R}_8^2} \rc 
\eer
where
\be
T_{(p,q)}=\frac{1}{2\pi \ls} \sqrt{p^2+\frac{q^2}{(g_s^{0B})^2}}
\ee

this is exactly the bosonic spectrum of type B theories compactified in a 
circle. The fermions make 
the difference because their ``$p$'' tensions can also be half integers.
This is what breaks the $SL(2,Z)$ symmetry of type IIB, but not completely 
because there are still bound states of 0B an IIB strings. When one takes the 
coupling to be very small, one has to keep only the $(\frac{1}{2},0)$-strings.
The spectrum of the $(p,q)$-strings depends on whether $p$ is integer or not.
If it is, fermions are periodic around the world-sheet and the spectrum is 
that of the IIB strings while if $p$ is half an odd number, the multiplets are
those of type 0B strings.

\begin{figure}
%%Begin InstantTeX Picture
\let\picnaturalsize=N
\def\picsize{3in}
\def\picfilename{m.eps}
%If you do not have the picture file add:
%\let\nopictures=Y
%to the beginning of the file.
\ifx\nopictures Y\else{\ifx\epsfloaded Y\else\input epsf \fi
\let\epsfloaded=Y
\centerline{\ifx\picnaturalsize N\epsfxsize \picsize\fi
\epsfbox{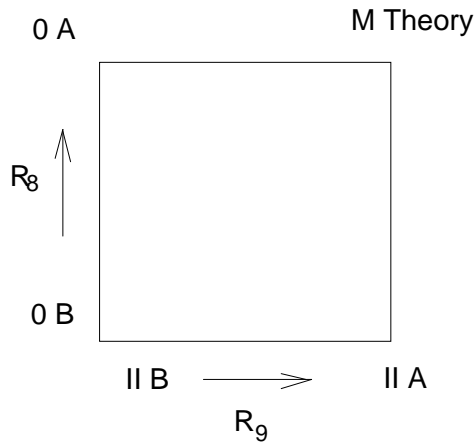}}}\fi
%%End InstantTex Picture
\caption{Square of theories related by dualities in the orbifolded torus.}
\label{fig1}
\end{figure}

  This also shows that, as proposed in \cite{gab}, the tachyon acquires a mass
when
the coupling is not small. The reason is that when $g_s^{0B}$ is large, one should 
make a flip (an S-duality) and consider $R_8$ to be the quantum circle. The 
Hamiltonian is the same again but the fermions are periodic in the 
world-sheet, and therefore generate the IIB supersymmetric spectrum. 
The first
level is the massless one so we conclude that, at least at large coupling,
the theory is sensible and does not contain tachyons. The swap of the role of
the two circles depends on their sizes and determines the phase structure of
M theory in this compactification (see fig.\ref{fig1}).

   When one takes the coupling constant to zero and keeps only the 
$(\frac{1}{2},0)$-strings, one can calculate their partition function. As in
the supersymmetric case \cite{sfm}, it is necessary to integrate over the
dual of the photon field after adding a topological contribution
\be
\epsilon^{\mu\nu\lambda}F_{\mu\nu}\partial_\lambda \phi^8
\ee
With this new scalar $\phi$, the $SO(8)$ symmetry is restored. Also, the
fermions change their chirality as is usual in a T-duality, which is, 
ultimately, what we are doing. With this, it is immediate to see that the 
partition function of these strings is exactly the same as that of the 0A 
computed above. The only difference is the chirality of fermions. As a result,
this model reproduces type 0B string theory.

   The type IIB limit is the opposite ($g_s^{IIB}=R_8/R_9\rightarrow 0$). 
The type IIA string can be achieved sending now $R_8$ to infinity. With
a dimensional reduction of the $\sigma_8$ direction, the Hamiltonian simply
reduces to that of ref. \cite{dvv}. Fig.\ref{fig1} sketches all the 
dualities on this torus.

\section{Non-perturbative states}

 In usual supersymmetric matrix string theories one can study non-perturbative
objects finding out the fluxes of fields that contribute to the central charges
of the supersymmetry algebra. In this case we do not have that guide. However,
similar fluxes of fields can be seen to act as Dirichlet objects and therefore
they are unstable non-perturbative effects with renormalised masses and 
charges. In fact, in the derivation of their existence and their behaviour as
walls where the open strings can end \cite{bfm,i,dvv} only the bosonic sector 
is needed, so one can extrapolate known results to this case.

   We have already seen one of the non-perturbative objects, the D-string. 
Among others, the usual IIB massless spectrum has appeared. From the point 
of view of the 0B corner ($g_s=R_9/R_8$), they have acquired a large mass
due to the fact that they are strongly coupled, but still keep their form
as a supersymmetric multiplet, including the fermions.

  Another interesting case is that of D0-branes in 0A theory. The flux 
related to them, as well as to their bound states is

\be
\frac{1}{2 \pi} \oint d\sigma E_9=n
\ee
where $E_9$ is the electric field and $n$ is the charge of the bound state. 
As argued in \cite{i}, the presence of a constant electric field forbids the
other fields from carrying momentum so that there cannot be any excited 
oscillators
in the Yang-Mills and the object with that flux has to be point-like. What
makes it act as a Dirichlet point is precisely that fact: when a twisted
state is formed between a perturbative string and a D0-brane (that is, between
two matrices with and without constant background electric fields), there 
cannot be
a flux of momentum (oscillator number) across the part of the
world-sheet with constant electric field (D-particle) so that it acts as a 
wall against which the momentum of the string rebounds; that is why it 
provides it with Dirichlet boundary conditions.

  Membranes in 0A an D3-branes in 0B theories are related to non-commutative
configurations of fields like
\be
\Phi=\frac{i}{4 \pi}\int d\sigma_8 tr < \lc X^7,X^6 \rc >=T_2 A .
\ee

  In the 0B case one also has to integrate over the $\sigma_9$ direction, 
which is the third world-volume dimension of the D3-brane. The other branes
are related to other fluxes exactly like in ref.\cite{bfm,dvv}.

\section{Interactions}

  As in ref.\cite{dvv,motl}, interactions appear when the condition of 
infinite coupling in the Yang-Mills theory is relaxed a bit. With that 
assumption there are special points in the moduli space where part of the 
broken gauge symmetry is restored. This happens whenever at least two 
eigenvalues of the matrices coincide for one value of $\sigma_9$. The symmetry
is enhanced to $SU(2)$ and asymptotically the effect is the interchange of 
the eigenvalues, that is, the action of an element of the unbroken centre of 
the group, $Z_N$. This means that also in this case the interactions are local 
in the world-sheet and appear in the target space as splitting and joining 
processes. In the absence of any supersymmetry it is hard to find the 
operators in the CFT that act as vertices. That would involve dealing with a 
non-perturbative expansion of a Yang-Mills theory without any supersymmetry.
Also, some relation should be found between the running of the coupling
constant in the Yang-Mills and the string theory. 
The only thing we can say, for the moment, is that, in the infrared, they 
can be given a geometrical interpretation as string interactions.

\section{Conclusions}

   The model presented here is a simple orbifold construction of the matrix
model of strings that describes closed type 0 strings as infrared fixed
points of one and two dimensional Yang-Mills theories. The world-sheet action
is not supersymmetric but it contains fermions that, together with bosons
generate the correct spectrum. The partition function is modular invariant.

   The effect of the tachyon is to provoke an infrared divergence both in the 
partition function and in the long-range supergravity potential.

   The two-dimensional Yang-Mills theory describes the IIB/0B dual system
with $(p,q)$ bound states that can be purely bosonic or fill up a 
supersymmetric multiplet. In particular IIB or 0B strings appear as 
non-perturbative excitations of the dual theory. As a conclusion, the 0B 
theory, although completely bosonic perturbatively, does contain fermions
when the coupling is not zero. Also, this model contains, through dimensional
reductions, the two other corners of the duality, the type A theories.

   Other branes can also be described as fluxes. Without supersymmetry, these
objects are not protected against decay or renormalization of their masses and
charges, nevertheless they are true excitations of the model. A more profound
study would need a better knowledge of open type 0 strings.

\section*{Acknowledgements}

  I have to thank M. Laucelli Meana and D. Mateos for introducing me to the
topic and for conversations. I also thank E. \'Alvarez for discussions and
M. A. R. Osorio for carefully reading the draft. Finally I acknowledge the
hospitality of CERN, where this work was completed, and specially the students 
of the Theory Division, that make it a nice place to work and live.


\newpage
\begin{thebibliography}{99}

\bibitem{gab}
O.~Bergman and M.R.~Gaberdiel,
``Dualities of type 0 strings,''
hep-th/9906055.
%%CITATION = HEP-TH 9906055;%%

\bibitem{blum}
J.D.~Blum and K.R.~Dienes,
``Strong / weak coupling duality relations for non-supersymmetric string
                  theories,''
Nucl. Phys. {\bf B516}, 83 (1998)
hep-th/9707160.
%%CITATION = NUPHA,B516,83;%%

\bibitem{tdual}
Y.~Imamura,
``Branes in type 0 / type II duality,''
hep-th/9906090.
%%CITATION = HEP-TH 9906090;%%



\bibitem{var}
L.J.~Dixon and J.A.~Harvey,
``String theories in ten dimensions without space-time supersymmetry,''
Nucl. Phys. {\bf B274}, 93 (1986);
%%CITATION = NUPHA,B274,93;%%

O.~Bergman and M.R.~Gaberdiel,
``A non-supersymmetric open string theory and S duality,''
Nucl. Phys. {\bf B499}, 183 (1997)
hep-th/9701137;
%%CITATION = NUPHA,B499,183;%%

L.~Alvarez-Gaume, P.~Ginsparg, G.~Moore and C.~Vafa,
``An O(16) X O(16) Heterotic String,''
Phys. Lett. {\bf B171}, 155 (1986);
%%CITATION = PHLTA,B171,155;%%

N.~Seiberg and E.~Witten,
``Spin structures in string theory,''
Nucl. Phys. {\bf B276}, 272 (1986).
%%CITATION = NUPHA,B276,272;%%

\bibitem{rohm}
R.~Rohm,
``Spontaneous supersymmetry breaking in supersymmetric string theories,''
Nucl. Phys. {\bf B237}, 553 (1984).
%%CITATION = NUPHA,B237,553;%%

\bibitem{tse}
A.A.~Tseytlin and K.~Zarembo,
``Effective potential in non-supersymmetric SU(N) x SU(N) gauge theory and
                  interactions of type 0 D3-branes,''
Phys. Lett. {\bf B457}, 77 (1999)
hep-th/9902095.
%%CITATION = PHLTA,B457,77;%%



\bibitem{bfss}
T.~Banks, W.~Fischler, S.H.~Shenker and L.~Susskind,
``M theory as a matrix model: a conjecture,''
Phys. Rev. {\bf D55}, 5112 (1997)
hep-th/9610043.
%%CITATION = PHRVA,D55,5112;%%

\bibitem{taylor}
W.I.~Taylor,
``D-brane field theory on compact spaces,''
Phys. Lett. {\bf B394} (1997) 283
hep-th/9611042.
%%CITATION = PHLTA,B394,283;%%



\bibitem{dvv}
R.~Dijkgraaf, E.~Verlinde and H.~Verlinde,
``Matrix string theory,"
Nucl. Phys. {\bf B500}, 43 (1997)
hep-th/9703030.
\bibitem{motl}
L.~Motl,
``Proposals on non-perturbative superstring interactions,''
hep-th/9701025.
%%CITATION = HEP-TH 9701025;%%

\bibitem{sfm}
T.~Banks and N.~Seiberg,
``Strings from matrices,''
Nucl. Phys. {\bf B497}, 41 (1997)
hep-th/9702187.
%%CITATION = NUPHA,B497,41;%%


\bibitem{i}
J.~Puente Pe\~nalba,
``Non-perturbative thermodynamics in matrix string theory,''
hep-th/9904094.
%%CITATION = HEP-TH 9904094;%%;

\bibitem{bfm}
T.~Banks, N.~Seiberg and S.~Shenker,
``Branes from matrices,''
Nucl. Phys. {\bf B490} (1997) 91
hep-th/9612157.
%%CITATION = NUPHA,B490,91;%%


\end{thebibliography}
\end{document}